\documentclass[letterpaper,sort&compress]{aipproc}
\layoutstyle{6x9}
\usepackage{amsmath}

\newcommand{\minipar}[3][c]{%
  \begin{minipage}[#1]{#2}%
    \ifthenelse{\equal{#1}{t}}{\vspace{0pt}}{}%
    {#3}%
    \ifthenelse{\equal{#1}{b}}{\par\vspace{0pt}}{}%
  \end{minipage}%
}
\newcommand{\Et}[1][]{E_T^{#1}}
\newcommand{\Ftwobb}{F_2^{{b\bar{b}}}}
\newcommand{\Ftwocc}{F_2^{{c\bar{c}}}}
\newcommand{\GeV}{\,\text{Ge}\eVdist\text{V\/}}
\newcommand{\eVdist}{\kern-0.06667em}
\newcommand{\ee}{e^+e^-}
\newcommand{\figsub}[3]{\raisebox{#2}{\makebox[0pt][r]{(#3)\rule{#1}{0pt}}}}
\newcommand{\fig}[1]{Fig.~\ref{fig:#1}}
\newcommand{\frag}{D_{H/h}}
\newcommand{\ig}{\includegraphics}
\newcommand{\muf}{\mu_F}
\newcommand{\mur}{\mu_R}
\newcommand{\pQCD}{\hat{\sigma}_{Vi\to hX}}
\newcommand{\pbi}{\,\text{pb}^{-1}}
\newcommand{\pdf}{\phi_{i/p}}
\newcommand{\php}{\ensuremath{\gamma p}}
\newcommand{\pt}[1][]{p_T^{#1}}

\begin{document}
\title{Heavy Flavors in High Energy $ep$ Collisions}
\author{Meng Wang\\
  {on behalf of the H1 and ZEUS collaborations}}{
  address={Bonn University, Institute of Physics,
    Nu\ss allee 12, 53115 Bonn, Germany}
  , email={wangm@physik.uni-bonn.de}
}
\date{\today}
\begin{abstract}
  Most recent measurements of open charm and beauty production in high
  energy $ep$ collisions at HERA are reviewed. The measurements explored
  the different aspects of quantum chromo\-dynamics involved in the
  process of heavy flavor production.  The results are compared with
  perturbative theoretical calculations at next-to-leading order.
\end{abstract}
\maketitle

\section{Introduction}
The masses of heavy quarks, charm and beauty, provide hard scales for
perturbative quantum chromodynamics (QCD) calculations.  Measurements of
heavy flavor production therefore have been and continue to be of great
interest as a rich testing ground for the reliability of perturbative
QCD predictions.

HERA, which collides electrons or positron of energy $27.5\GeV$ with
protons of energy $920\GeV$ (or $820\GeV$ before 1998) resulting in a
center-of-mass energy of $318\GeV$ (or $300\GeV$), can test heavy flavor
production in a unique way. Two collider experiments, H1 and ZEUS, have
accumulated approximately $135\pbi$ of integrated luminosity by the end
of 2000.  After a major upgrade during 2001 and 2002, HERA is running at
much higher luminosity with the polarized electron beam and is referred
to as HERA~II for distinction.

\begin{figure}[tbp]
  \centering
  \ig[width=.36\textwidth]{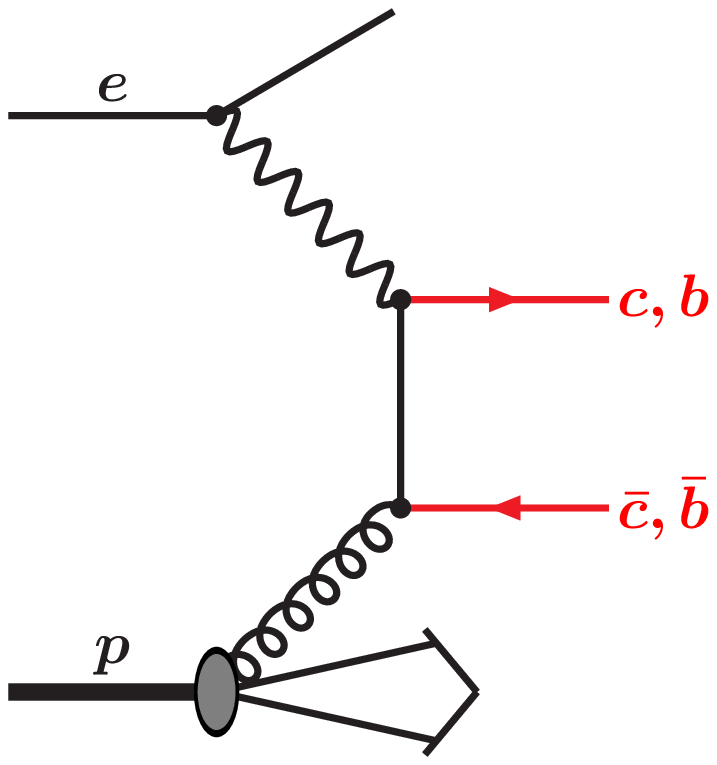}
  \caption{Heavy flavor production via Boson-gluon-fusion in the $ep$
    collision.}
  \label{fig:bgf}
\end{figure}

Heavy flavor production at HERA is dominated by boson-gluon-fusion
(BGF), as shown in \fig{bgf}. When the virtuality $Q^2$ of the exchanged
boson, mainly photon, is very small, $Q^2 \ll 1\GeV^2$, the virtual
photon resembles a real one and the collisions are referred to as $\php$
or photoproduction, in which a certain fraction of the photons can be
resolved with parton contents.  For large $Q^2$, the collisions are
called deep inelastic scattering (DIS). Only the most recent
measurements are presented here.

\section{Theory and Models}
In perturbative QCD (pQCD), single-particle inclusive cross
sections in $ep$ collisions can be factorized in the form
\[
\sigma_{ep\to HX} = \sum_i \pdf \otimes \pQCD \otimes \frag,
\]
where $\otimes$ denotes convolution. The sum is over all relevant
partons $i$, while $\pdf$ is a parton density function (PDF) of the
initial proton, $\pQCD$ is the pQCD calculable cross section of hard
scatter, and $\frag$ is the fragmentation function. In resolved
photoproduction, an additional PDF of the photon enters the calculation.
These terms are evaluated at a renormalization scale $\mur$ as well as a
factorization scale $\muf$, which is often taken at the same value as
$\mur$. The PDFs and the fragmentation functions are non-perturbative
and must be determined either experimentally or taken from models, but
they are universal. In heavy flavor production, the mass of the heavy
quark provides an additional hard scale hence yielding more accurate
calculations.

Up to next-to-leading order (NLO), the fixed-flavor-number scheme (FFNS)
or ``massive'' scheme from~\citet{Frixione:1995qc} for photoproduction
and from~\citet{Harris:1997zq} for DIS and the zero-mass
variable-flavor-number scheme (ZMVFNS) or ``massless'' scheme
from~\citet{Cacciari:1997du} and from~\citet{Heinrich:2004kj} are
compared with the data.  They differ in the treatment of mass of heavy
quark. In the massive scheme, the heavy quarks are non-active flavors in
the proton and are produced through hard scatter such as BGF, while in
the massless scheme, the heavy quarks are just the contents of the
proton and can enter the reaction directly. The massive scheme works
well near the threshold of heavy quark production and the massless
scheme works better in the higher kinematic region. A scheme matching
the two also exists but is not used for the measurements presented here.

Monte Carlo (MC) models, based on leading order QCD and parton shower
approaches, are used for acceptance calculations in all measurements,
and sometimes as alternative QCD predictions. They include
PYTHIA~\cite{Sjostrand:2000wi} and HERWIG~\cite{Corcella:2000bw} which
differ in the treatment of parton showers and heavy quark fragmentation.

\section{Open Charm Production}
\begin{figure}[tbp]
  \centering
  \minipar{.51\textwidth}{%
    \ig[width=\textwidth]{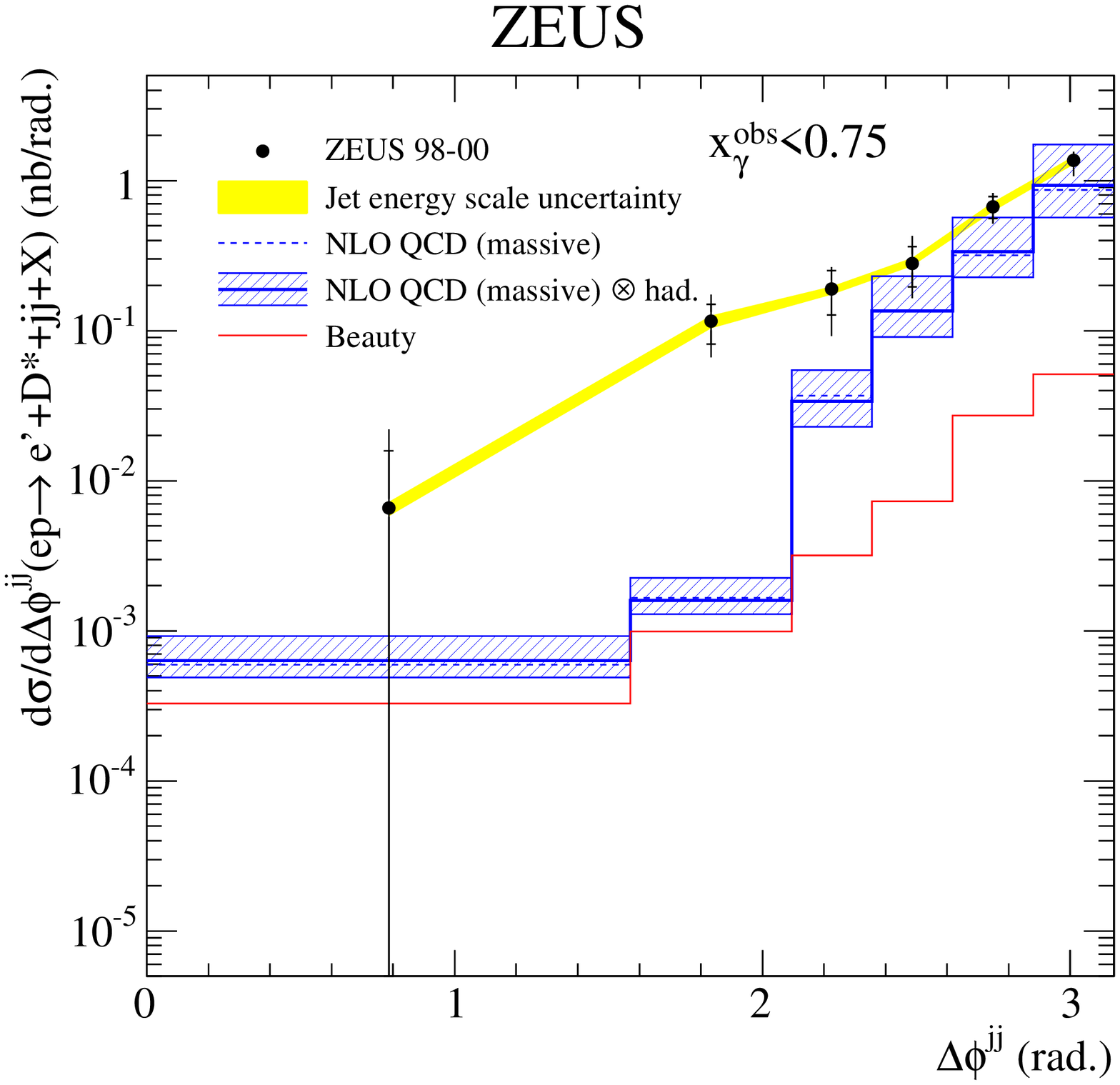}%
    \figsub{2em}{3em}{a}
  }%
  \minipar{.49\textwidth}{%
    \ig[width=\textwidth]{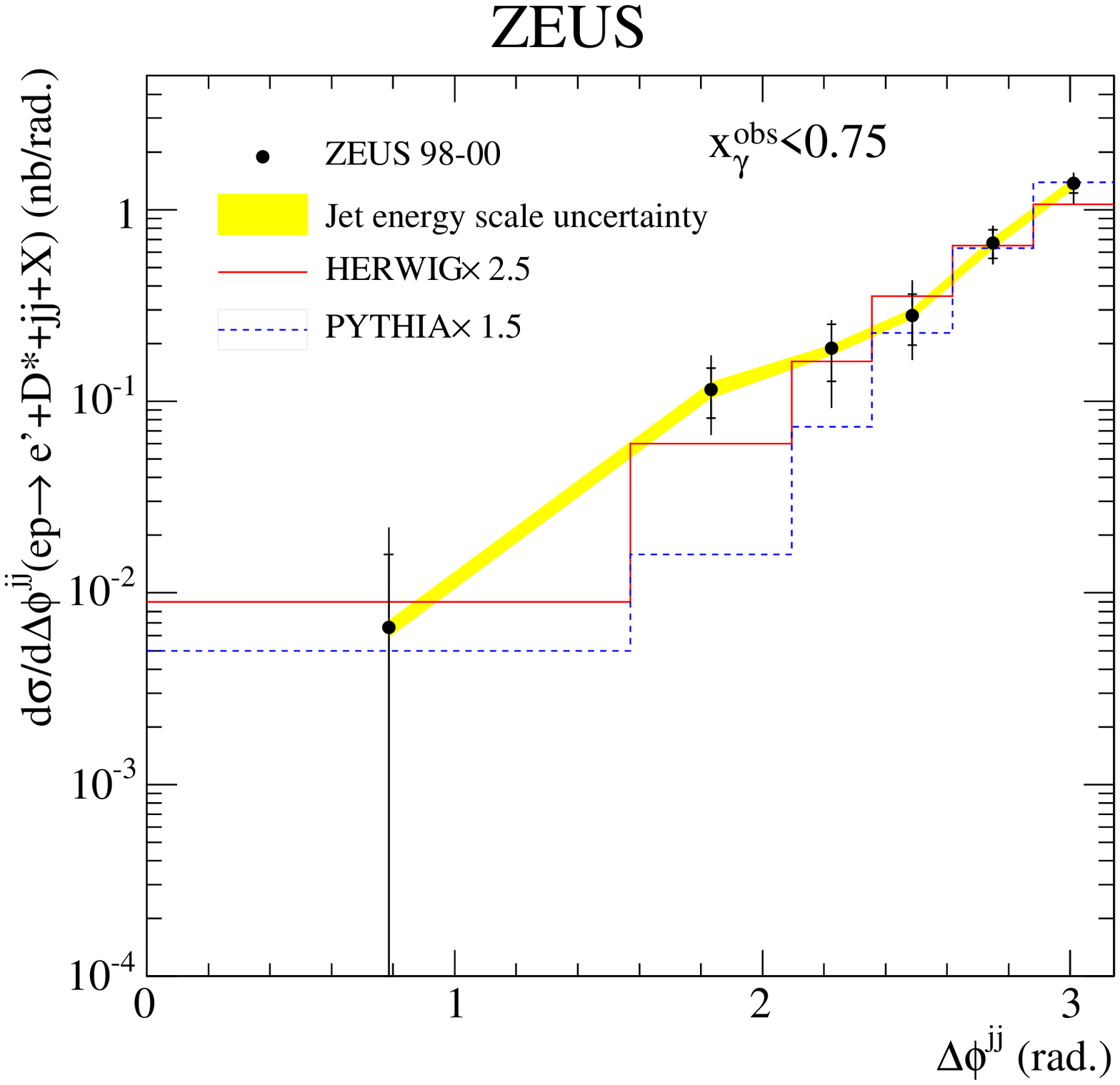}%
    \figsub{2em}{3em}{b}
  }
  \caption{Azimuthal angular difference of the two highest $\Et$ jets in
    $D^*$ photoproduction for the resolved enriched sample, compared to
    (a) a NLO QCD calculation and (b) Monte Carlo models.}
  \label{fig:dijet}
\end{figure}

Inclusive jet cross sections and dijet correlations in $D^*$
photoproduction have been studied by ZEUS~\cite{zeus:2005zg}. NLO
predictions in the massive and massless schemes show reasonable
agreement with the data in all inclusive jet cross sections as well as
some dijet correlation cross sections, but the massive prediction shows
a large deviation at low $\Delta\phi^{jj}$, azimuthal angular difference
of two jets, and high $(p_T^{jj})^2$, squared transverse momentum of
dijet system. The discrepancy is enhanced in the resolved-enriched
sample, as shown in \fig{dijet}a. However, the HERWIG MC model describes
the shape of the data well, \fig{dijet}b. This indicates the necessity
of higher-order calculations or additional parton showers in current NLO
calculations.

\begin{figure}[tbp]
  \centering
  \minipar[b]{.52\textwidth}{%
    \ig[width=\textwidth]{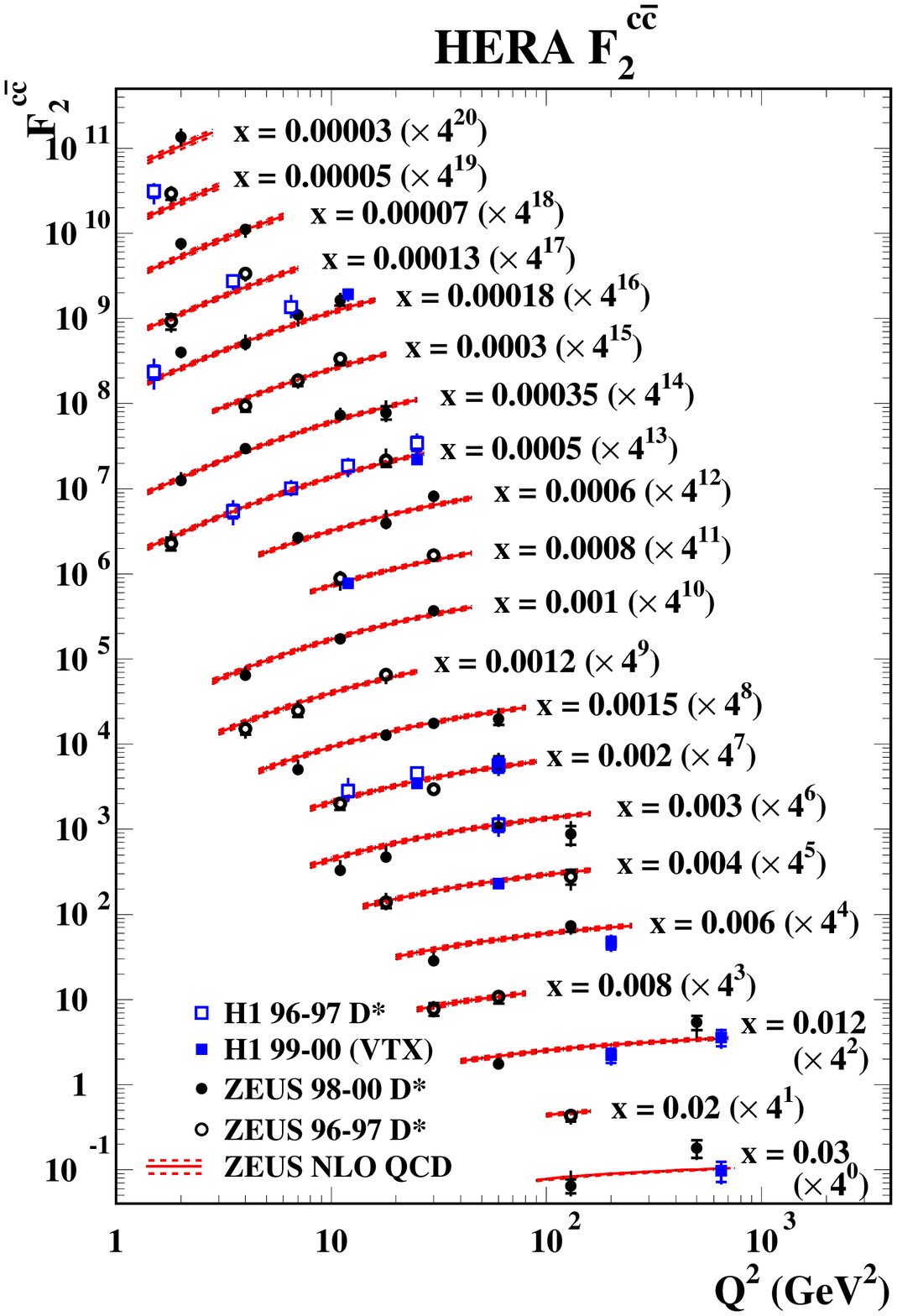}%
    \figsub{1em}{1.3\textwidth}{a}
  }%
  \hspace{4ex}%
  \minipar[b]{.41\textwidth}{%
    \ig[width=\textwidth]{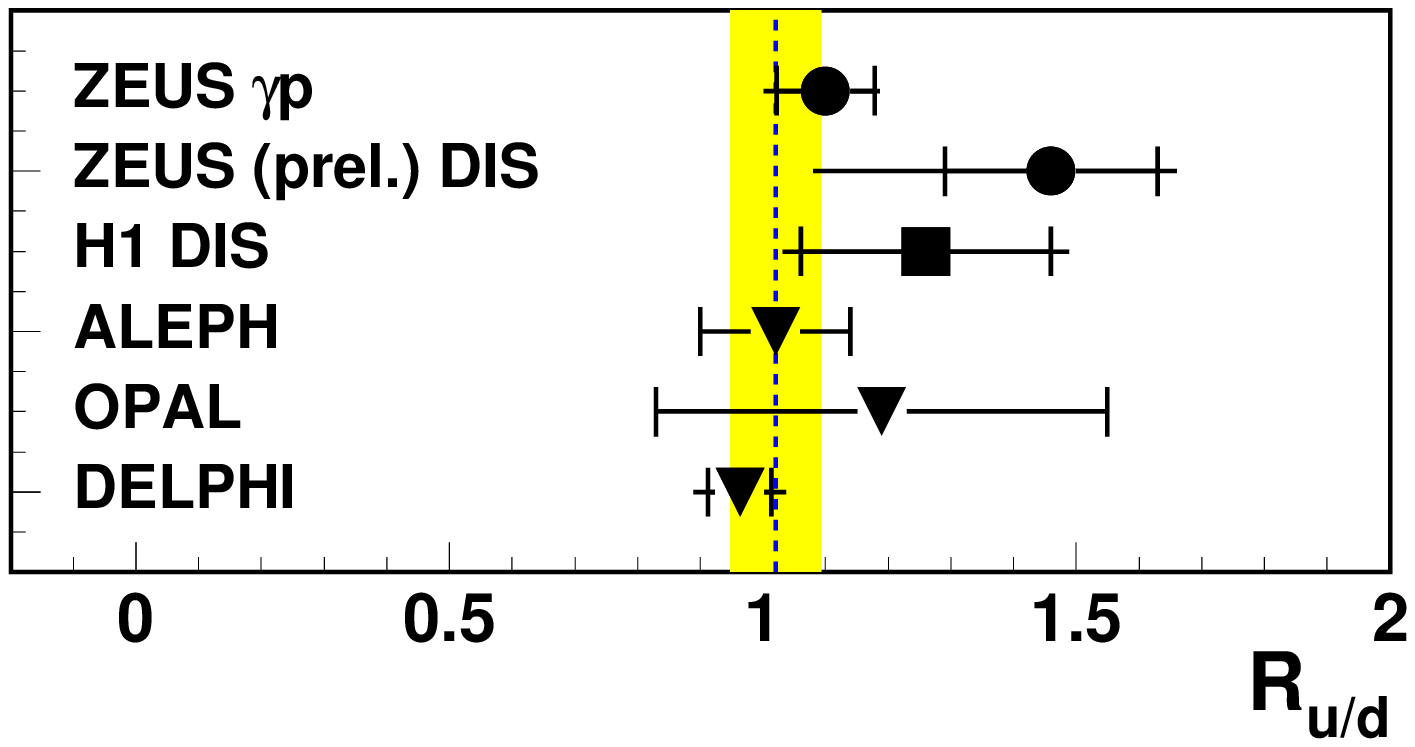}%
    \figsub{.8em}{2.1em}{b}
    \ig[width=.99\textwidth]{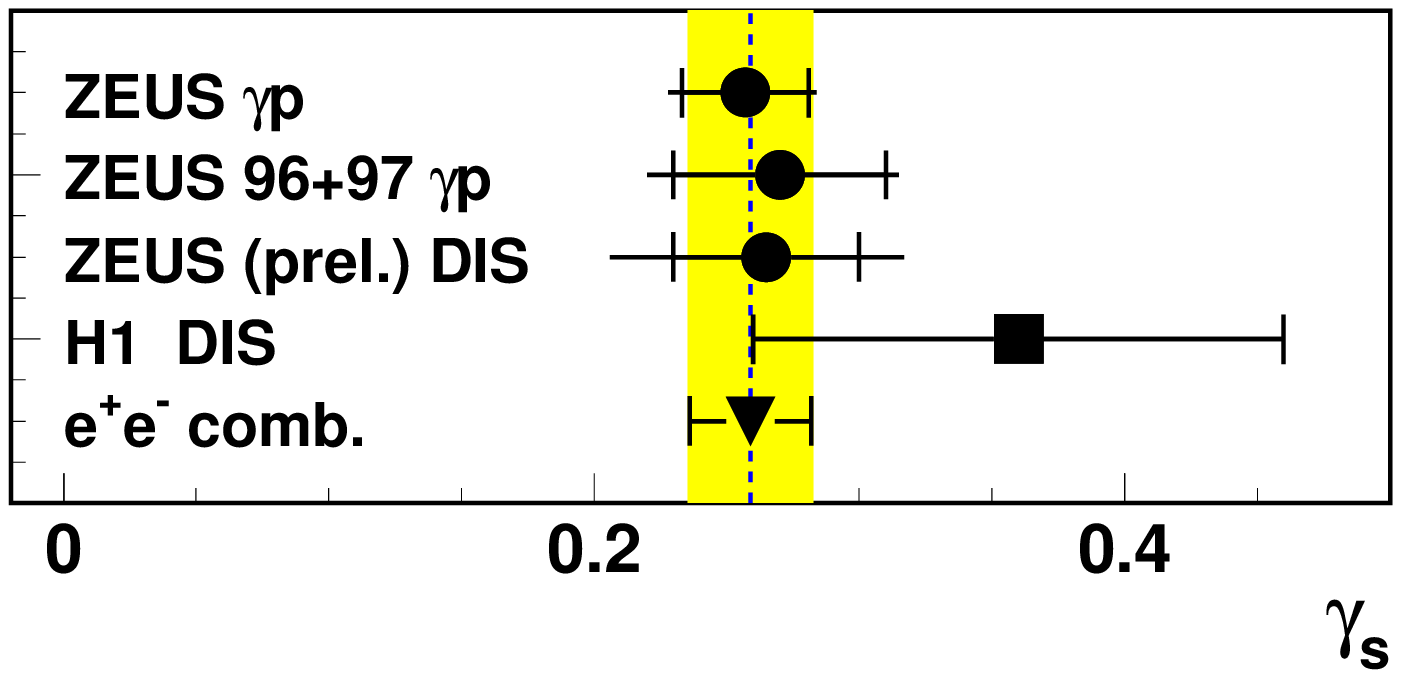}%
    \hfill\figsub{.8em}{2.1em}{c}
    \ig[width=1.02\textwidth]{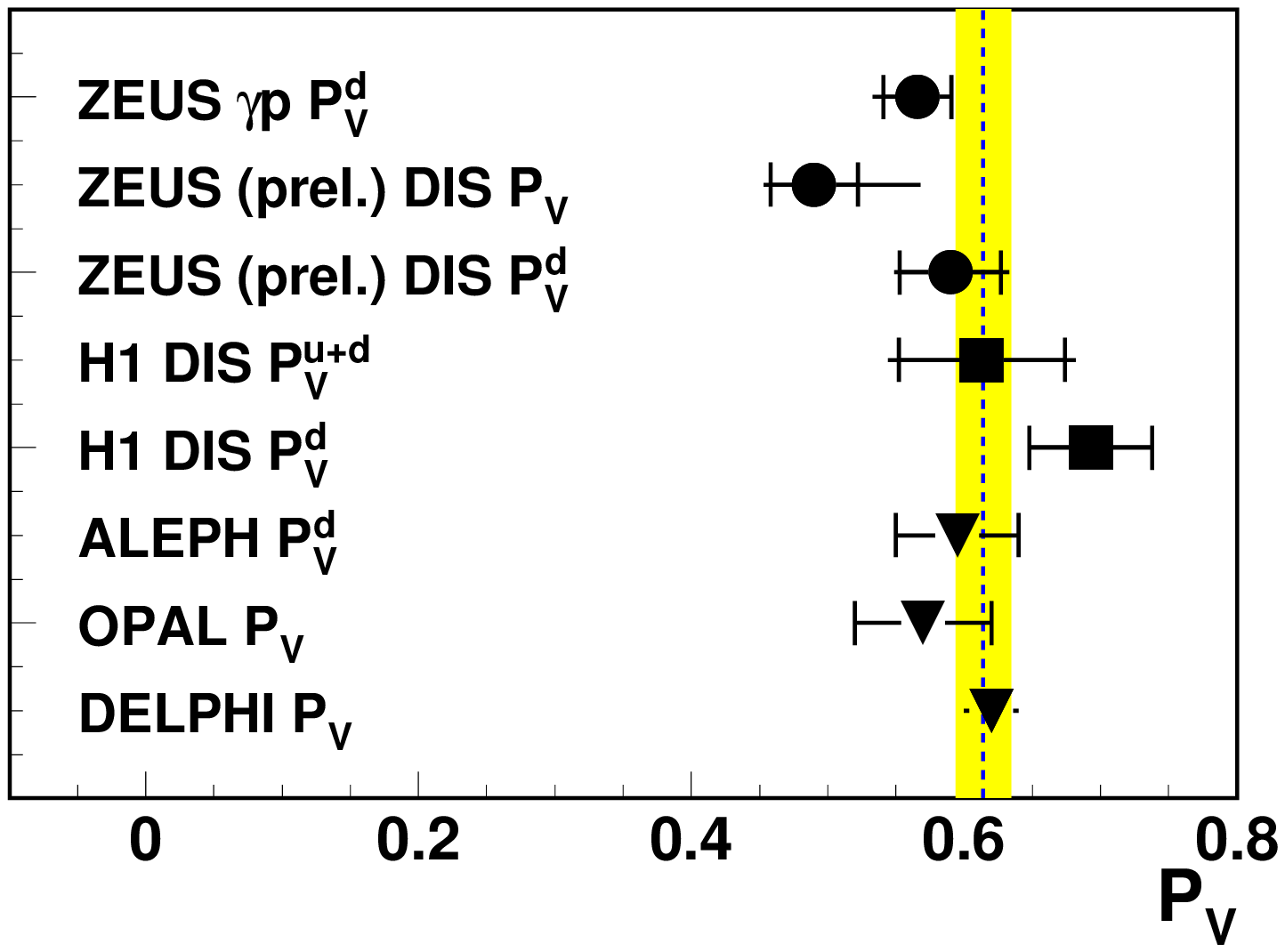}%
    \figsub{.9em}{2.1em}{d}
  }
  \caption{(a) $\Ftwocc$ as a function of $Q^2$ for different values of
    the Bjorken scaling variable $x$; \mbox{(b-d) Comparisons} of charm
    fragmentation ratios $R_{u/d}$, $\gamma_s$ and $P_V$. (Comparisons
    of charm fragmentation fractions, $f(c\to D, \Lambda)$, are not
    shown for brevity.)}
  \label{fig:Ftwocc}
\end{figure}

Open charm production in DIS is directly sensitive to the gluon contents
in the proton.
Measurements~\cite{zeus:1999ad,h1:2001zj,zeus:2003rb,h1:2004az,h1:2005iw}
of the charm contribution to the proton structure function, $\Ftwocc$,
are shown in \fig{Ftwocc}a.  The results are well described by a QCD
prediction based on a QCD fit of the inclusive structure function
$F_2$.  A measurement exploring the transition region between DIS and
photoproduction has also been reported recently~\cite{lp2005:265} as
well as the first charm measurement from HERA~II data~\cite{lp2005:271}.

Charm fragmentation is traditionally measured at $\ee$ experiments and
the results are then adapted into MC simulations or theoretical
calculations. Recently, H1 and ZEUS made measurements on charm
fragmentation ratios $R_{u/d}$ (the ratio of neutral to charged
$D$-meson rates), $\gamma_s$ (the strangeness suppression factor) and
$P_V$ (the fraction of $D$-meson in a vector state) as well as
fragmentation fractions, $f(c\to D,
\Lambda)$~\cite{h1:2004ka,zeus:2005mm,lp2005:266}. The results are
compared to those of $\ee$, \fig{Ftwocc}b-d, and confirm the
universality of fragmentation.  Measurements of the fragmentation
function have also been made by H1~\cite{lp2005:407} and
ZEUS~\cite{ichep2002:778}, and are in agreement with universality. too.

\section{Open beauty production}
\begin{figure}[tbp]
  \centering
  \minipar[t]{.45\textwidth}{%
    \ig[width=\textwidth]{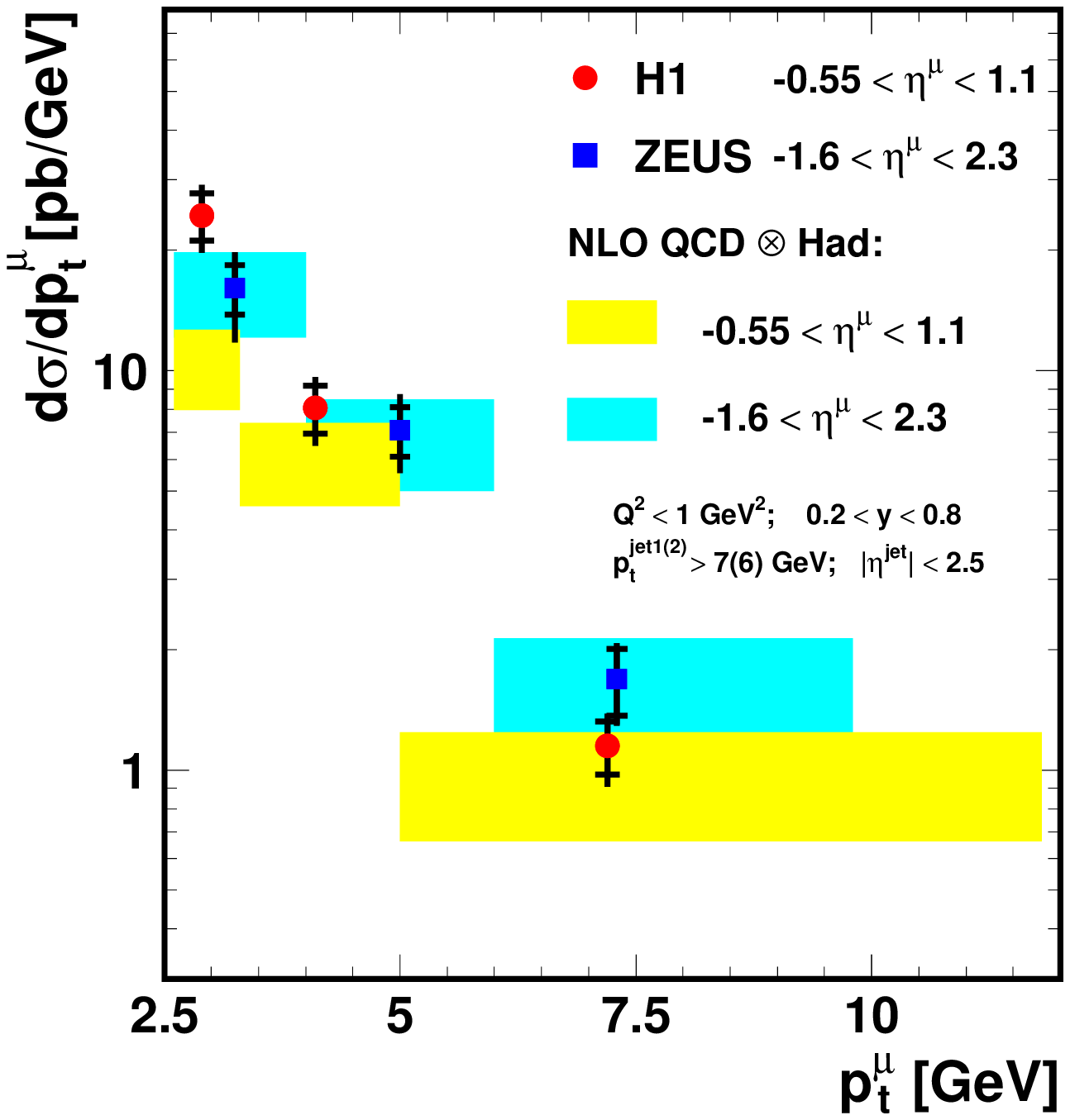}%
    \figsub{.7\textwidth}{3em}{a}
  }%
  \hspace{2em}%
  \minipar[t]{.45\textwidth}{%
    \ig[width=\textwidth]{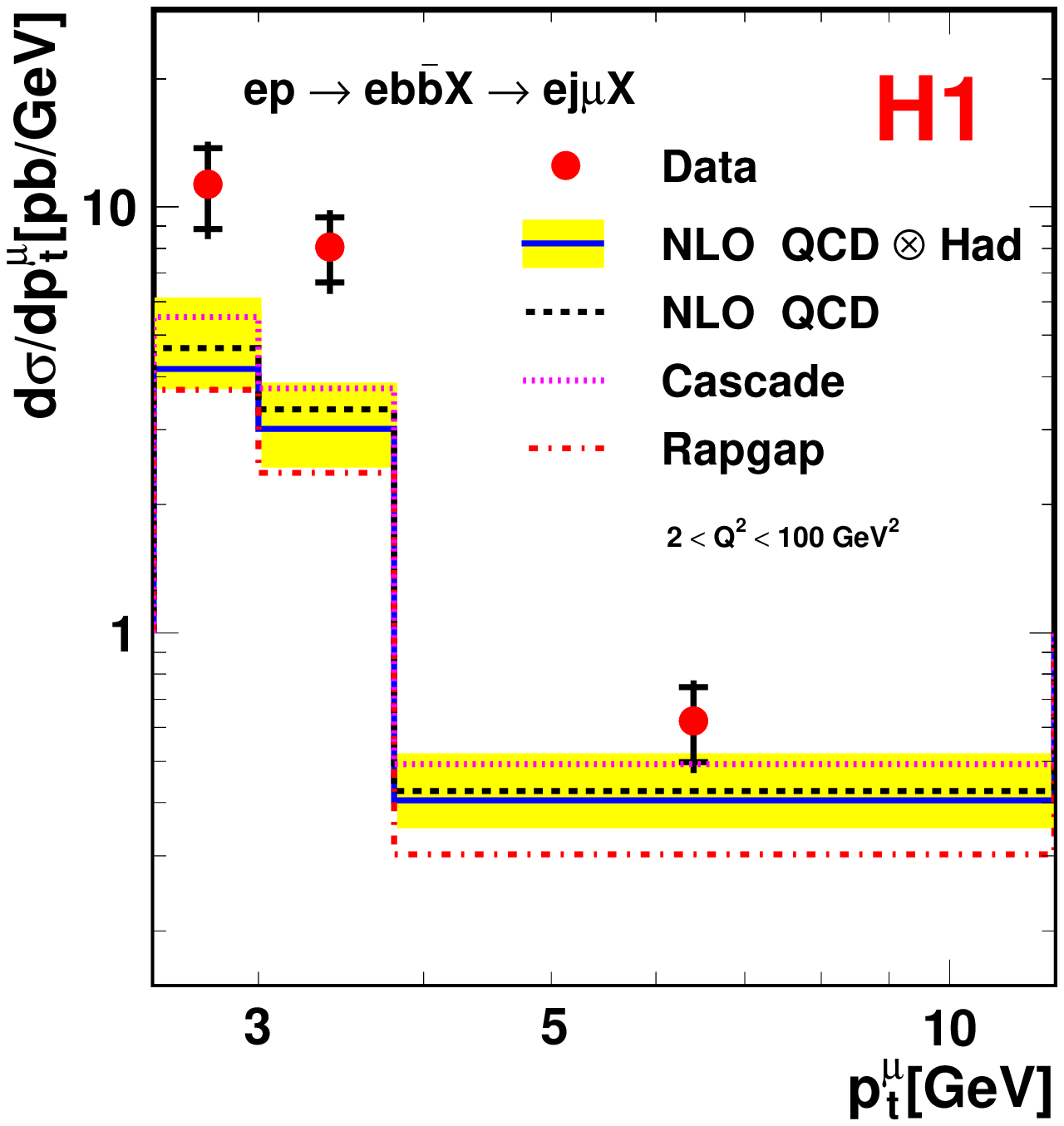}%
    \figsub{.7\textwidth}{3em}{b}
  }
  \caption{Open beauty production as a function of $\pt[\mu]$ for
    (a) dijet photoproduction from the H1 and ZEUS experiments and (b)
    inclusive jet DIS from the H1 experiment. (The measurement from the
    ZEUS experiment for (b) has the similar result and is not shown for
    brevity.)}
  \label{fig:mujet}
\end{figure}

Because of the larger mass, the pQCD calculation for open beauty
production should be more reliable than for open charm. Some of the
previous measurements~\cite{h1:1999nr,zeus:2000nz} have found the
prediction to be below the data by up to a factor 3. Recent measurements
by H1 and ZEUS required events containing jets in addition to high $\pt$
muons, which are traditionally used to tag beauty
quarks~\cite{zeus:2004xy,zeus:2004tk,h1:2005zc}. As shown in
\fig{mujet}a, the measurements in photoproduction agree very well
between the two collaborations and are well described by the NLO
calculations although the H1 data is above the calculation at low
$\pt[\mu]$. In DIS, similar agreement exists, although discrepancies at
low $\pt[\mu]$ as well as forward $\eta^\mu$ are observed by both
collaborations, shown in \fig{mujet}b as an example.

\begin{figure}[tbp]
  \centering%
  \ig[height=.53\textheight]{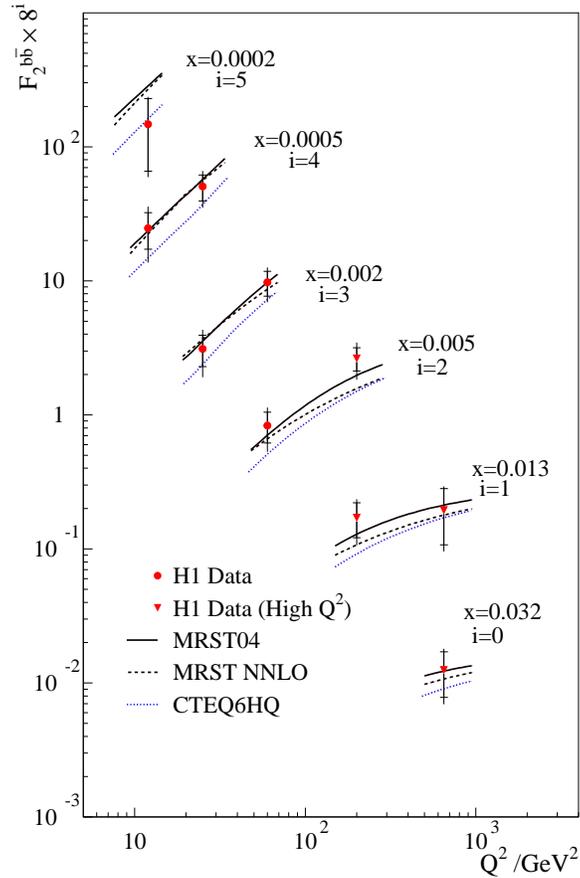}
  \caption{$\Ftwobb$ as a function of $Q^2$ for different $x$.}
  \label{fig:Ftwobb}
\end{figure}

New measurements using inclusive impact parameters of tracks from decays
of long lived charm and beauty hadrons have been performed in DIS by the
H1 collaboration~\cite{h1:2004az,h1:2005iw}. They are the first
measurements of $\Ftwobb$, \fig{Ftwobb}. The data are well described by
the NLO predictions and a recent calculation using NNLO structure
functions~\cite{Thorne:2005nz}.  The values of $\Ftwocc$ also confirm
the previous data and have been put in \fig{Ftwocc}a. 
Significantly improved further measurements are expected from the
increased HERA~II statistics and the corresponding detector improvements.
First preliminary results on beauty in photoproduction using the new
ZEUS micro-vertex detector have already been obtained~\cite{hep2005:359}.

\section{Conclusions}
\renewcommand{\baselinestretch}{.97}

At HERA, measurements on heavy flavor production are getting more
extensive and precise. New results improve the understanding of the
heavy quark contributions to the proton structure function as well as
the hadronic behavior of the photon. The universality of the
fragmentation has also been confirmed by the recent measurements. The
perturbative QCD calculation generally describes the data at
next-to-leading order although discrepancies still exist in some
threshold regions. Therefore higher order corrections beyond the
existing models and calculations might be needed. More exciting and
accurate measurements are definitely expected in the future from HERA~II
with higher luminosity and improved detector ability of the H1 and ZEUS
experiments.

\section{Acknowledgments}
I really appreciate the invitation from the conference organizers. I am
also grateful to A. Geiser and M. Wing for productive discussions. This
work is supported by the German Federal Ministry of Education and
Research (BMBF) under Contract No. HZ4PDA.

\bibliography{}
\bibliographystyle{aipproc}
\end{document}